# A Security Protocol for Multi-User Authentication

*Srikanth Chava*

**Abstract**: *In this note we propose an encryption communication protocol which also provides database security. For the encryption of the data communication we use a transformation similar to the Cubic Public-key transformation [1]. This method represents a many-to-one mapping which increases the complexity for any brute force attack. Some interesting properties of the transformation are also included which are basic in the authentication protocol.*

## Introduction

The globally distributed database networks today demand secure communication protocol between the users. Techniques of public key cryptography use one-to-one mapping while authenticating the users. The many-to-one mapping may also be used for authentication of multiple users, which is what we propose here.

The general transformation used in this paper is $C = M^X$ modulo N, where N is a product of two individual primes, say P and Q, and X divides the Euler's totient function $\varphi(N)$, which is a generalization of the cubic public key transformation [1]. Being a many-to-one mapping this can help in deceiving the hacker. The transformation can be used for authenticating multiple users and its properties can help us defining some protocol which allows us to provide a secure communication and database.

## Properties of the transformation

In the transformation $C = M^X$ modulo N, Cipher C will have X roots and these will satisfy properties that are basic in creating our security protocol. N should be selected in such a way that $\varphi(N)/X$ is a positive integer.

*Relational properties:*
*Property (1):* Different roots of a cipher C have their difference as multiple of either of the two factors of N.

*Property (2):* If two of the roots for a cipher are given the rest of the values can be found by a simple function:
Let one of the roots multiply the other by factor F then $F \times M(a)$ modulo $N = M(b)$.
Where $M(a)$=one of the roots and $M(b)$=one of the roots and not equal to $M(a)$

Inverse property:
*Property (3):* If all the roots of a cipher C are multiplied and taken modulo of N we automatically get back the cipher C
$C = M^x$ modulo N

$(M_1 \times M_2 \ldots M_x)$ modulo $N = C$.

This property follows from the fundamental theorem of algebra.

Each cipher C has X roots and these will satisfy properties that can be exploited in creating our security protocol. If we are given one root we can easily find out the other roots and we also can find the actual cipher. It means that if there are X number of users and they all can send different messages using similar cipher which can actually deceive the intruder. They can find out who the authenticated user is by using the properties (1) and (2) as they clearly show the relation between the roots obtained from the tags attached to their ciphers.

**Example:**

$C = M^5$ modulo N
Let, $N = 55 = 5 \times 11 = P \times Q$
$\varphi(N)/X = 40/5 = 8$.

For C=23
$F(m) = (3, 23, 38, 48, 53)$,

(I) Property (1):
$m_4 - m_2$
$48 - 23 = 25 = 5 \times 5 = 5 \times P$.

(II) Property (2):
$m_1 \times F = m_4$
$3 \times 16 = 48$
$16 \times 23 = 368$ modulo $55 = 38 = m_3$
Similarly, we can find all the roots.

(III) Property (3):
$(m_1 \times m_2 \times m_3 \times m_4 \times m_5)$ modulo $N = C$

$(3 \times 23 \times 38 \times 48 \times 53)$ modulo $55 = 23 = C$.

| | | | |
|---|---|---|---|
| 1 | 1 | 28 | 43 |
| 2 | 32 | 29 | 54 |
| 3 | 23 | 30 | 10 |
| 4 | 34 | 31 | 1 |
| 5 | 45 | 32 | 32 |
| 6 | 21 | 33 | 33 |
| 7 | 32 | 34 | 34 |
| 8 | 43 | 35 | 10 |
| 9 | 34 | 36 | 1 |
| 10 | 10 | 37 | 12 |
| 11 | 11 | 38 | 23 |
| 12 | 12 | 39 | 54 |
| 13 | 43 | 40 | 10 |
| 14 | 34 | 41 | 21 |
| 15 | 45 | 42 | 12 |
| 16 | 1 | 43 | 43 |
| 17 | 32 | 44 | 44 |
| 18 | 43 | 45 | 45 |
| 19 | 54 | 46 | 21 |
| 20 | 45 | 47 | 12 |
| 21 | 21 | 48 | 23 |
| 22 | 22 | 49 | 34 |
| 23 | 23 | 50 | 10 |
| 24 | 54 | 51 | 21 |
| 25 | 45 | 52 | 32 |
| 26 | 1 | 53 | 23 |
| 27 | 12 | 54 | 54 |

**Table 1: $C = M^5$ modulo 55**

**Table 2: f (m)**

| C1=1  | 1 | 16 | 26 | 31 | 36 | f(m)=(1,16,26,31,36) |
|-------|---|----|----|----|----|----------------------|
| C2=23 | 3 | 23 | 38 | 48 | 53 | f(m)=(3,23,38,48,53) |
| C3=32 | 2 | 7  | 17 | 32 | 52 | f(m)= (2,7,17,32,52) |
| C4=34 | 4 | 9  | 14 | 34 | 49 | f(m)= (4,9,14,34,49) |
| C5=45 | 5 | 15 | 20 | 25 | 45 | F(m)=(5,15,20,25,45) |

**Method**

The proposed network protocol can have X number of users; here X comes from the general transformation $C = M^x$ modulo N. There can be any number of users and X can be a very large number also. Unlike the one to one mapping in the existing public key cryptography the general transformation provides us with multiple authentications for multiple users. Here the user is provided with a user interface which actually stores the algorithm and performs all the calculations that are necessary for the authentication which is similar to the concept of secret hardware public key cryptography [2]. Each user sends a value $M_X$ through the cipher C and a tag $T_x$ and the number of users in the protocol decides the value of X. Access is provided to the users and the database server with the key and if once he is authenticated he sends the cipher C and also the value of X which acts as a public key [4],[5]. For authentication of the database server, two users must come together to get the actual encrypted file. The important part of this model is that the encrypted database is provided security with the help of key combination [4].

The users transmit cipher C along with a tag information T where $T_X = (M_X-C$ modulo $P)/P$, where P is the private key and it is smallest of the two factors of the public key N [1]. The user on other end gets back his message by simply following the inverse equation:

$M_X = T_X \times P + C$ modulo P.

If the root is proved to be one of the roots from the same cipher the user is authenticated and access is granted.

The tag information $T_x$ is also sent along with the cipher C and can be separated from combination as the other user knows the cipher C that is matched to the root $M_x$.

## Inter User Authentication

Authentication between the users is to make sure that the user is a genuine member of the users and this is done by following the first property proposed above. Let User1 wants access with the other users say User2 and User3, initially they should provide their key through the tag attached to the cipher C and they are verified by the user interface simultaneously and after that he can actually communicate the messages in encrypted form with both the users at the same time[3]. The decryption is very simple and it follows from the inverse equation proposed. The applications for this inter user authentication include requesting the access for the database with authentication and requesting the access for the files which the user doesn't have any access.

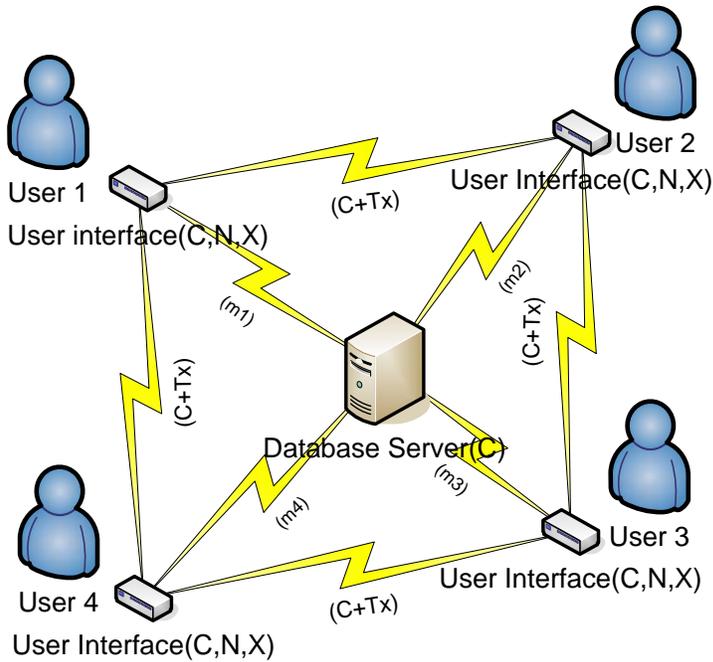

**Figure 1: Network Protocol**

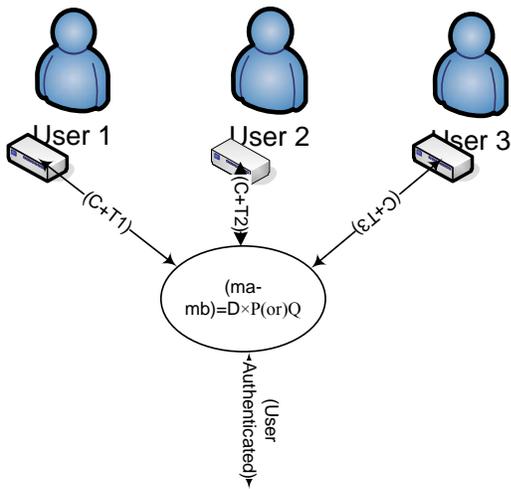

**Figure 2:** User Authentication.

## Intra User Authentication

Intra User Authentication of a user is used when he accesses the database server. Here we are proposing a different technique to grant the users the access to the database in a unique way. The files are stored in a database which is encrypted and according to the protocol a user needs any one combination from other users to access the encrypted files from the database [4]. He requests the access in the Inter User communication protocol and when he is authenticated he can get one of the users to access the database server and then if they provide their individual keys it is possible to get the encrypted file from the properties 2 and 3 that the proposed above[5]. In this way if user wants to access the database server he has to be authenticated twice which provides high security. After getting the encrypted file if any other users want the file they can request them using the Inter User communication protocol discussed above and the user who has the file can actually broadcast it to the multiple users at the same time.

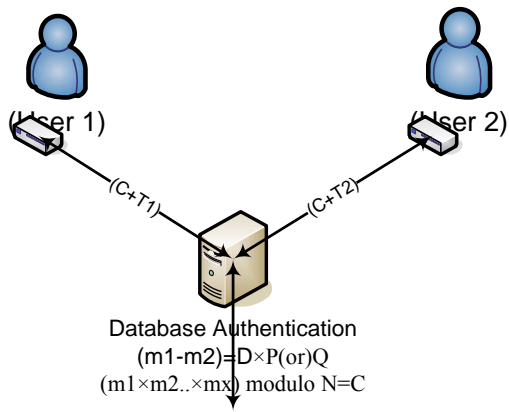

**Figure 3:** Database Authentication.

## Summary and Conclusion

The transformation which we proposed used in this paper can provide high level of security since there are X messages for one cipher. This increases the level of complexity, thus providing high level of security in authenticating the users in a protocol. Moreover the messages or roots to the cipher always increase with the increase in the value of exponent. It opens a new door to the multiple user public key cryptography which is better than the existing public key system [3].

The database is also made secure by our method as the data in the database is encrypted into one cipher which is a representation of many messages but looks like the same because of the properties of the transformation we have used. The access for the database server is provided only after authenticated the users twice. However we use the secret hardware system here so the transformation used should always be unknown. The decryption as described needs that the users come together to decrypt the cipher, in this process the users can share the private keys so it is suggested that the value of N is refreshed periodically.

The properties of the transformation provided in this paper may be exploited in many ways. This generalized transformation may be further used to make the authentication more secure for multiple user applications.